\documentclass[journal]{IEEEtran}
\usepackage[utf8]{inputenc}

\usepackage{lipsum}
\usepackage{url}

\ifCLASSINFOpdf
\else
\fi

\usepackage{graphicx} 
\usepackage{listings}
\usepackage{xcolor}

\definecolor{codegreen}{rgb}{0,0.6,0}
\definecolor{codegray}{rgb}{0.5,0.5,0.5}
\definecolor{codepurple}{rgb}{0.58,0,0.82}
\definecolor{backcolour}{rgb}{0.95,0.95,0.92}

\lstdefinestyle{mystyle}{
    backgroundcolor=\color{backcolour},   
    commentstyle=\color{codegreen},
    keywordstyle=\color{magenta},
    numberstyle=\tiny\color{codegray},
    stringstyle=\color{codepurple},
    basicstyle=\ttfamily\footnotesize,
    breakatwhitespace=false,         
    breaklines=true,                 
    captionpos=b,                    
    keepspaces=true,                 
    numbers=left,                    
    numbersep=5pt,                  
    showspaces=false,                
    showstringspaces=false,
    showtabs=false,                  
    tabsize=2
}

\usepackage[english]{babel}
\usepackage[utf8]{inputenc}
\usepackage{amsmath}
\usepackage{amsfonts}
\usepackage{graphicx}

\usepackage{comment}
\usepackage[colorinlistoftodos]{todonotes}
\usepackage{algorithm}
\usepackage{algpseudocode}
\usepackage{cool,siunitx}
\Style{LogShowBase=always}
\usepackage{mathtools}

\usepackage{fourier, erewhon}
 \usepackage{caption}
\usepackage{graphicx}
\usepackage{subcaption}
\DeclareCaptionSubType*[Alph]{table}
\DeclareCaptionLabelFormat{mystyle}{Table~\bothIfFirst{#1}{ ̃}#2}
\usepackage{diagbox}
\usepackage{subcaption,siunitx}

\usepackage{multirow}
\usepackage{collcell}

\usepackage[english]{babel}
\usepackage[utf8]{inputenc}
\usepackage{amsmath}
\usepackage{amsfonts}
\usepackage{graphicx}
\usepackage[colorinlistoftodos]{todonotes}
\usepackage{algorithm}
\usepackage{algpseudocode}
\usepackage{amssymb}
\usepackage{amsthm}

\usepackage{amsmath}
\usepackage{pifont}
\usepackage{amsmath}

\usepackage{color}

\usepackage{amsmath,amssymb}
\usepackage[english]{babel}
\usepackage{amsmath}
\usepackage{amsfonts}

\usepackage{pifont}

\usepackage{algorithm}  
\usepackage{algpseudocode}  
\usepackage{amsmath}  
\ifCLASSINFOpdf

\else

\fi
\usepackage{blindtext}

\IEEEoverridecommandlockouts
\makeatletter

\def\footnoterule{\kern-3\p@\hrule \@width 2in \kern 2.6\p@}
\makeatother

\begin{document}
%
\title{RAG-Based Fuzzing of Cross-Architecture Compilers}
%
%
%

\author{
    \IEEEauthorblockN{Rana Elnaggar, Brian Delgado, and Jason M. Fung}\\
    \IEEEauthorblockA{Intel Corporation\\}
    E-mail: rana.elnaggar@intel.com, brian.delgado@intel.com, and jason.m.fung@intel.com
    
}

\maketitle

\begin{abstract}
OneAPI is an open  standard that supports cross-architecture software development with minimal effort from
developers. It brings DPC++ and C++ compilers which need to be thoroughly tested to
verify their correctness, reliability, and security. Compilers have numerous code flows and optimization
features. This process requires developers with deep understanding of the different compiler flows to
craft testcases specific to target paths in the compiler. This testcase creation is a time-consuming and costly process. In this
paper, we propose a large-language model (LLM)-based compiler fuzzing tool that integrates the concept of
retrieval-augmented generation (RAG). This tool automates the testcase generation task and relieves experienced
compiler developers from investing time to craft testcase generation patterns. We test our proposed approach on the Intel
DPC++/C++ compiler. This compiler compiles SYCL code and allows developers to offload it to different architectures, e.g. GPUs and CPUs from different vendors. Using this tool, we managed to identify 87 SYCL code test cases that lead
to output value mismatch or compiler runtime errors when compiled using Intel DPC++ and clang++ compilers
and run on different architectures. The testcases and the identified unexpected behaviors of the compilers under test were obtained within
only few hours with no prior background on the compiler passes under tests. This tool facilitates efficient
compiler fuzzing with reduced developer time requirements via the dynamic testcase creation capability provided by an LLM with RAG.
\end{abstract}


%
\IEEEpeerreviewmaketitle

\section{Introduction}
The Intel DPC++ compiler enables the same source code to be compiled to generate binaries of different vendor architectures. Accomplishing this purpose requires numerous complicated flows
that are implemented across several development teams. Thus, this compiler needs to be thoroughly tested to
ensure that it does not crash, nor introduce exploitable bugs in the generated binaries or differing behavior on other architectures. Compiler fuzzing has been widely adopted to test
different compilers (e.g., LLVM, Nvidia NVCC and OpenCL compilers) \cite{WhiteFox_oopsla}. Prior work has relied on three major
paradigms for compiler fuzzing: (1) closed-box fuzzing, which generates random code snippets based on the
target language grammar and features of the target language; (2) partially-closed box fuzzing, which utilizes
information about test cases coverage to guide further test generation; and (3) open-box fuzzing which
generates test cases based on inspecting the source code of the compiler passes. Closed-box and partially
closed box fuzzing techniques suffer from low compiler code coverage and the inability to target specific
compiler passes. The ability to test a specific compiler pass is crucial during the compiler development phase.
Thus, when closed-box and partially closed box fuzzing techniques are applied, multiple bugs remain
unrevealed by fuzzing campaigns that continue to run over years \cite{yang2011finding}. Open-box fuzzing addresses the
limitations in closed and partially-closed fuzzing, however, it utilizes hand-written code generation templates
and policies to cover language features related to a specific compiler pass \cite{livinskii2020random}. Thus, this methodology is highly
labor-intensive. It requires experienced compiler developers with extensive knowledge in compiler features and
compiler passes to develop these templates and generation policies. In addition, a custom template needs to be
implemented for every compiler pass under test. This severely affects the scalability and feasibility of open-box
fuzzing.

\begin{figure*}[t]
    \centering
\includegraphics[width=0.9\textwidth]{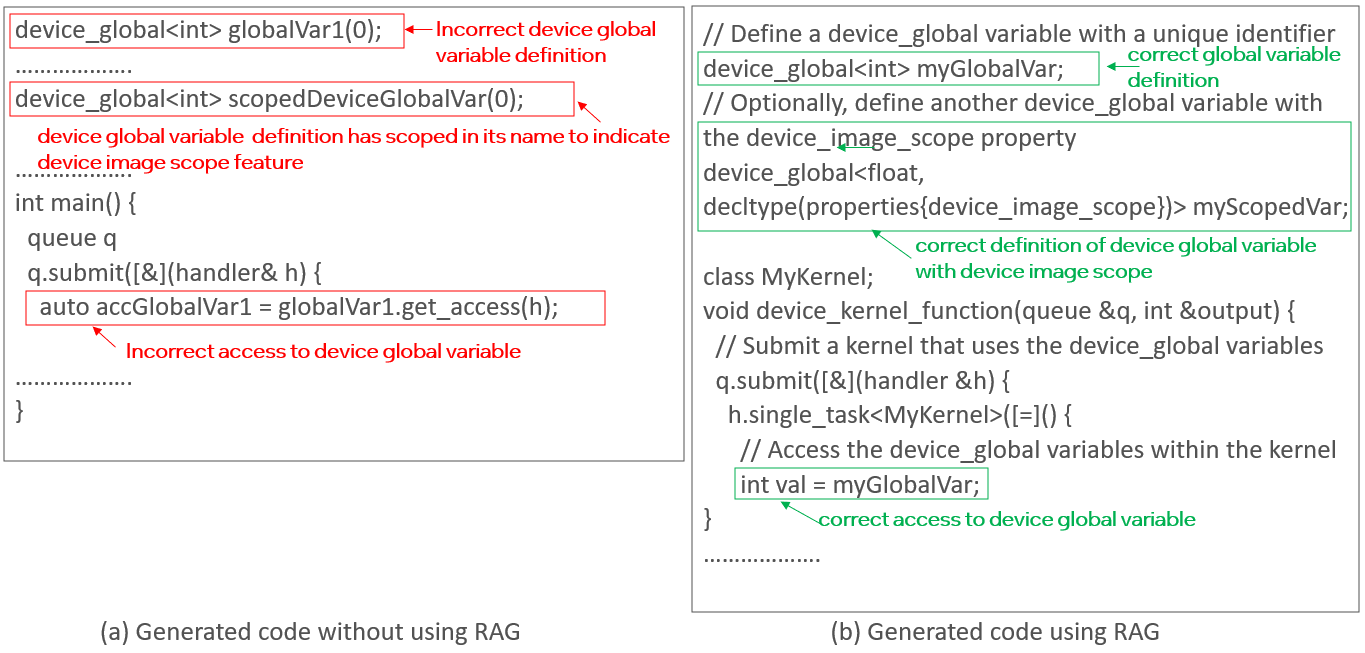}
\vspace{-0.1cm}
\caption{Motivation behind using RAG. (a) sample generated code without using RAG; (b) Sample generated code without using RAG }
\label{motivation_RAG_}
\vspace{-0.5cm}
\end{figure*}

Recently, large language models (LLMs) have proven their success in code generation. The work in \cite{WhiteFox_oopsla} proposes
to generate test cases using LLMs with few-shot in-context learning instead of expert-defined code generation templates and policies. They
manually select functions within specific compiler passes and then provide the function implementation to
LLMs to generate source code to test the target function within the compiler flow. This approach showed superior
results on LLVM and deep learning compilers fuzzing. However, when we applied their methodology to the Intel DPC++ compiler without few-shot learning, we
have observed that the test cases that are generated do not comply with SYCL language features and suffer from hallucination \cite{10.1145/3571730} and made up code as shown in Fig.~\ref{motivation_RAG_}.  Thus, this approach did not achieve our goal of a fuzzing tool that can be used by non-compiler experts.

The reason behind this observation is that there is plenty of information related to
LLVM compiler passes under test in \cite{WhiteFox_oopsla}. Thus, the LLM has been trained on various sources of data covering
LLVM optimizations under test. However, in the case of SYCL-specific compiler passes incorporated in Intel DPC++/C++ compiler, they are actively being developed. Thus, there is high possibility that the LLM does not
incorporate enough information about the Intel DPC++ compiler to generate correct code. Therefore, in this work we
propose to utilize the retrieval augmented generation (RAG) paradigm to provide LLMs with sufficient context that
aids it in code generation even when LLMs have not been explicitly trained on Intel DPC++ compiler features
and optimization flows. The proposed framework could easily be integrated by compiler engineering teams
with the Intel DPC++ compiler development environment to enable continuous fuzzing of the compiler passes as
they are being developed. 

The contributions of this work are listed below.
\begin{enumerate}
    \item This is the first paper to evaluate utilizing LLMs to fuzz test Intel DPC++ compilers. 
    \item We extend prior work \cite{WhiteFox_oopsla} by utilizing RAG methodology to generate syntactically and semantically correct sycl code to fuzz test Intel DPC++ compilers. 
    \item We present an end-to-end flow for RAG-based Intel DPC++ compiler fuzzing that can be extended to other compilers. 

\end{enumerate}

The rest of the paper is organized as follows. Section \ref{related_work} presents related work. Section \ref{e2e_tool} explains the proposed approach. Section \ref{exp_results} presents experimental setup and discusses the results. Section \ref{limitations_} outlines limitations and future work. Finally, Section \ref{conclusions_} concludes the paper.  
\section{Related Work}
\label{related_work}
\begin{figure}[t]
    \centering
\includegraphics[width=0.3\textwidth]{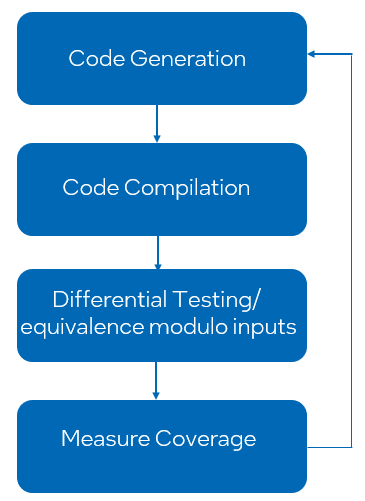}
\vspace{-0.1cm}
\caption{Typical compiler fuzzing flow.}
\label{compiler_fuzz_flow}
\vspace{-0.5cm}
\end{figure}
Compilers are significantly complex software. Fuzzing compilers has been an open research problems for over a decade. Over the years, various methodologies have been proposed to efficiently fuzz compilers. The typical flow for fuzzing compilers is as shown in Fig. \ref{compiler_fuzz_flow}. There is a code generation module to generate random code that triggers various compiler components. Differential testing is then performed to compile the generated code using different compiler versions, different compilers or different compiler flags and optimization flags. If the outcome of the execution of the generated binaries is different, a compiler bug is flagged. 
We classify the compiler fuzzing literature into two categories, the pre-Large language models (pre-LLM) era and post-LLM era.

In the pre-LLM era, random code is generated using language-grammar generators \cite{csmith_,clsmith_,cudasmith_,afl_fuzzer}
or hand-crafted templates by experienced compiler experts for each compiler component/transformation pass/optimization pass under tests \cite{yarpgen_,loop_optimization_}. In the post-LLM era, code is generated using LLMs \cite{WhiteFox_oopsla}.

\subsection{Pre-LLM compiler fuzzing} 
The work in \cite{csmith_} proposes a random test case generation methodology that targets fuzzing multiple C language compilers (e.g., GCC and LLVM, CIL, TCC, and Open64). The random generation of programs in this approach relies on a subset of C language grammar that includes various data types, structures and control flow statements. Undefined and unspecified behavior are avoided by a combination of static and run-time analysis of the generated test cases. The random test case generator has 40,000 lines of code, which makes it very complicated to maintain and extend. In addition, it does not cover all the C language features and specifications. The possible compiler bugs are detected using differential testing. In differential testing, the same source code is compiled with different optimization levels on different compilers and compiler versions. The compiler that generates a different answer compared to its peers is marked to have a possible issue. This methodology managed to report over 300 issues in GCC and LLVM compilers by running it over the span of three years. The detected issues include compiler crash errors where the compiler violates specific assertions or terminates unexpectedly before the generation of the source code binaries. They also include the generation of incorrect binaries that when executed at run-time generates unexpected behavior (e.g., incorrect output or unexpected termination).

The work in~\cite{clsmith_} proposes to fuzz opencl compilers. Bugs are detected in the compilers under test using differential testing and equivalence modulo inputs using multiple devices and OpenCL versions. The tool proposed in~\cite{clsmith_} extends the methodology proposed in~\cite{csmith_}. It randomly generates parallel OpenCL kernels with OpenCL features such as vectors, barriers, atomic sections, and atomic reduction modes. The work in~\cite{cudasmith_} adapts the tool proposed in~\cite{clsmith_} to fuzz Cuda compilers. 

Another approach, Nautilus~\cite{aschermann2019nautilus}, leverages a combination of fuzzer feedback along with grammar-based inputs. The approach provides inputs that pass syntactic and semantic checks to enhance effectiveness. The tool uses a provided grammar and stores a context free grammar tree-based representation of it. New fuzz permutations can be chosen by using selection rules from the tree and the observed coverage used to determine the usefulness of the input.


The work in~\cite{yarpgen_} randomly generates test cases via hand-written scripts to test various scenarios pertaining to loop optimizations. Undefined behavior is captured at the generation phase using static value checks and rewrite rules. The design of these handwritten optimization-specific scripts require extensive expertise in compiler details. In addition, new scripts need to be written to target new optimizations. Thus, this limits the feasibility of extending the proposed method to other compiler optimizations and transformation phases.

Compared to prior work, our proposed approach uses fewer lines of code which makes it simpler to maintain. In addition, it can be applied to any compiler that targets any programming language without triggering a change in the tool's source code. Also, covering all language features is feasible without the need to add extra lines of code to the implementation of the proposed tool.

\subsection{Post-LLM Era}
The work in \cite{WhiteFox_oopsla} proposes to use LLMs to generate random code to fuzz LLVM and deep learning compilers. This work requires that compiler developers define compiler transformation functions that they are interested in testing. Then the code of the transformation function is passed to an LLM to generate the characteristics that are needed to trigger the transformation function. Next, the extracted characteristics are passed to a code-generation LLM to generate test code. This work proposes to use few-shot in-context learning where examples for characteristics generation and test code generation are provided to the LLM as part of the context to generate new test cases. 

Compared to this work, we use zero-shot learning along with RAG in our proposed tool to reduce the hallucinations generated by the LLMs without having to use few-shot examples. 
\section{Proposed Work}
\label{e2e_arch}

\begin{figure*}[t]
    \centering
\includegraphics[width=0.9\textwidth]{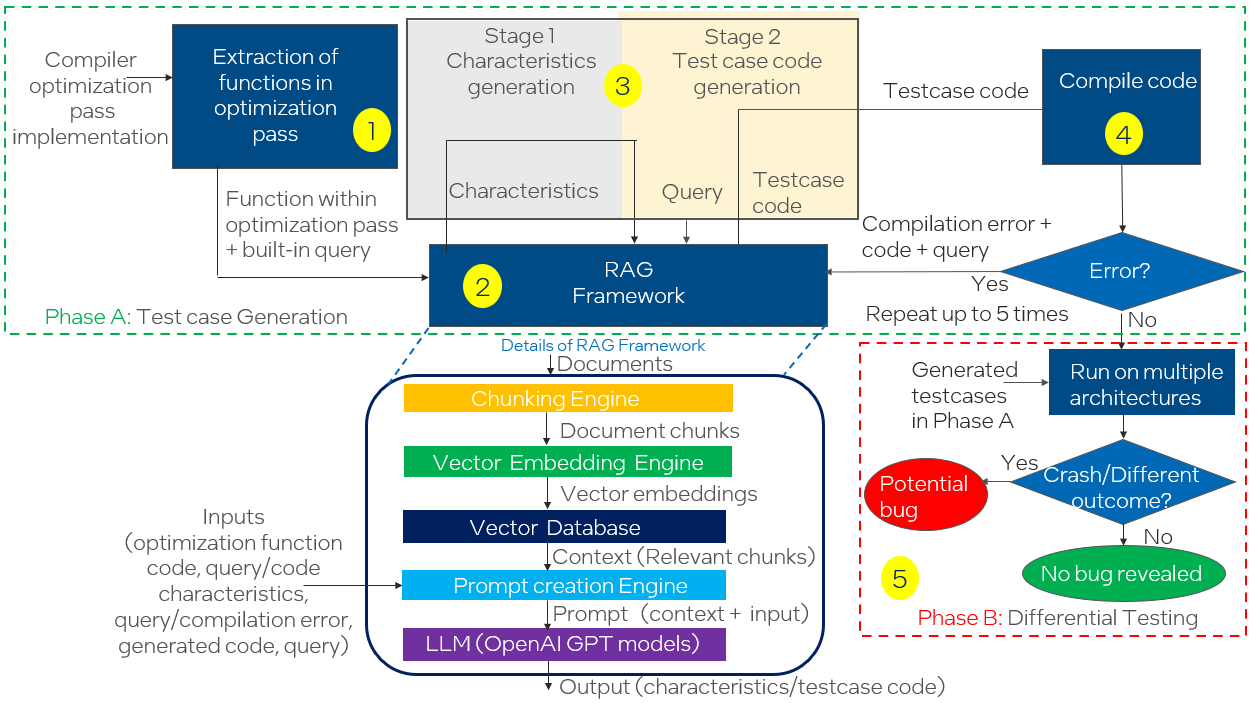}
\vspace{-0.1cm}
\caption{Proposed Architecture. The numbers 1-5 correspond to the tool modules described in Section \ref{e2e_arch}}
\label{e2e_tool}
\vspace{-0.5cm}
\end{figure*}

We propose an end-to-end tool which is illustrated in Fig. \ref{e2e_tool}. This end-to-end flow is based on the work in \cite{WhiteFox_oopsla}. It consists of 2 phases. Phase A takes as input (1) the compiler-related documents, SYCL specifications and the compiler source code to generate working source code snippets that are used to stress-test the compiler (output of the tool in Phase A). Phase B leverages the generated test code to perform differential testing across XPU architectures from various vendors to identify compiler issues. 

The proposed tool has the following
modules. \begin{enumerate}
    \item \textit{Extraction of functions in the compiler pass:} The input to this module is the code that
implements the compiler pass under test. This module extracts the code of each function in the compiler pass
under test. This function code would then be passed to a RAG framework to extract the characteristics of the
SYCL code that needs to be compiled to trigger this function in the compiler pass. The code of each function extracted by this module is passed along with a query to the RAG framework. 
\item \textit{RAG framework:} The basic
structure of the RAG framework is shown in Fig. \ref{e2e_tool}. RAG consists of a document chunking engine. It splits the
document into chunks. Each chunk can be formed of a single sentence, paragraph, or an entire section of the
document. The vector embedding of each chunk would be generated and the chunk text representation and the
embedding vector representation are stored in a vector database. Vector database is a database that can be
used to store vector representations~\cite{pg_vector}. It has efficient algorithms to retrieve data based on vector operations.
When RAG is used in inference, inputs are received by the RAG framework. The embedding of the input are
computed using the embedding generation engine. The distance between the chunk embedding stored in the
vector database and the embedding of the inputs to the RAG platform would be computed. If the computed
distance is below a specific threshold, the text related to the chunk embedding would be retrieved from the
vector database to form the text context that would be passed to the LLM. The output of the LLM would then be
passed to the subsequent stages~\cite{rag_}. In this work, we utilize Intel internal tool to implement the RAG
framework.
\item \textit{Test case generation module:} This module has two stages, (1) characteristics generation; and (2) test case
code generation. In the characteristics generation stage, the code implementation of each function in the compiler pass is passed along with a query to the RAG framework. The
query that is used in the characteristics generation phase is shown in Listing.~\ref{query_sycl1}. The RAG framework generates the
characteristics of the SYCL code needed to trigger the function under test. An example of a function in a compiler
pass, the generated characteristics, and the test code output are shown in listings \ref{lst:Example_},\ref{lst:Example2_}, and \ref{lst:Example3_}.

\begin{figure*}[t]
\begin{lstlisting}[frame=single,caption={Query that is used in the characteristics generation phase},label={query_sycl1}]
      You are an expert compiler developer. Please describe the characteristics of the device kernel implemented in SYCL that can trigger the optimization pass +{ Optimization Pass Name}+   in LLVM. The device kernel code characteristics should trigger all the lines in the following function/namespace/class WITHOUT EXPLICITLY CALLING THE FUNCTION ITSELF. The description should be concise and clear. Use code to illustrate patterns or constraints as needed. Please only describe the characteristics of the device kernel code. Do not describe the function/namespace/class code. MAKE SURE THAT THE REQUIREMENTS DONOT INCLUDE HAVING A FUNCTION OR METHOD THAT CALL ANY OTHER FUNCTION.
      > Code of the function/class/namespace
      {Code of function in optimization pass}

\end{lstlisting}
\end{figure*}

\begin{figure*}

\begin{lstlisting}[frame=single,caption={ An example of a function in DeviceGlobal compiler
pass~\cite{deviceg_}},label={lst:Example_}]
DeviceGlobalPropertyMapTy collectDeviceGlobalProperties(const Module &M) {
  DeviceGlobalPropertyMapTy DGM;
  auto DevGlobalNum = count_if(M.globals(), isDeviceGlobalVariable);
  if (DevGlobalNum == 0)
    return DGM;

  DGM.reserve(DevGlobalNum);

  for (auto &GV : M.globals()) {
    if (!isDeviceGlobalVariable(GV))
      continue;

    DGM[getGlobalVariableUniqueId(GV)] = {
        {{getUnderlyingTypeSize(GV), hasDeviceImageScopeProperty(GV)}}};
  }

  return DGM;
}
\end{lstlisting}
\end{figure*}

\begin{figure*}
\begin{lstlisting}[frame=single,caption={ Generated function characteristics},label={lst:Example2_}]
 To trigger the optimization pass `DeviceGlobals` in LLVM without explicitly calling the function `collectDeviceGlobalProperties`, 
the device kernel code in SYCL must exhibit certain characteristics that involve the use of `device_global` variables. Here are 
the characteristics that would trigger the optimization pass:

1. **Presence of Device Global Variables**: The device kernel code must declare and use one or more `device_global` variables. 
These are global variables that are recognized by the SYCL runtime as being accessible across both host and device.

2. **Unique Identification**: Each `device_global` variable must be uniquely identifiable. 
This is typically achieved through the use of a special attribute or naming convention that the LLVM pass can recognize.

3. **Underlying Type Size**: The size of the type of the `device_global` variable must be determinable at compile time. 
This is necessary for the `getUnderlyingTypeSize` function to retrieve the size information.

4. **Device Image Scope Property**: The `device_global` variable may optionally be declared with the `device_image_scope` property. 
This property indicates whether the variable is shared across all device images on a device or is specific to a single device image.

Here is an illustrative example of a SYCL device kernel code that would trigger the `DeviceGlobals` optimization pass:

```cpp
#include <CL/sycl.hpp>
using namespace sycl;

// Define a device_global variable with a unique identifier
device_global<int> myGlobalVar;

// Optionally, define another device_global variable with the device_image_scope property
device_global<float, decltype(properties{device_image_scope})> myScopedVar;

class MyKernel;

void device_kernel_function(queue &q) {
  // Submit a kernel that uses the device_global variables
  q.submit([&](handler &h) {
    h.single_task<MyKernel>([=]() {
      // Access the device_global variables within the kernel
      int val = myGlobalVar;
      float scopedVal = myScopedVar;
      // Perform computations...
    });
  });
}

int main() {
  queue q;
  device_kernel_function(q);
  return 0;
}
```

In this example, the presence of `myGlobalVar` and `myScopedVar` as `device_global` variables would be
detected by the LLVM pass when it processes the module containing this kernel. The pass would then
collect properties such as the size of the underlying type and whether the `device_image_scope` property is present.
The unique identification could be derived from the variable's name or an internal attribute applied by the SYCL compiler
during the compilation process.   
\end{lstlisting}
\end{figure*}

\begin{figure*}
\begin{lstlisting}[frame=single,caption={ Generated output test code},label={lst:Example3_}]
#include <sycl/sycl.hpp>
#include <array>
#include <iostream>
#include <string>
#if FPGA_HARDWARE || FPGA_EMULATOR || FPGA_SIMULATOR
#include <sycl/ext/intel/fpga_extensions.hpp>
#endif

using namespace sycl;
using namespace sycl::ext::oneapi::experimental;

// Define a device_global variable with a unique identifier
device_global<int> myGlobalVar;

// Optionally, define another device_global variable
with the device_image_scope property
device_global<float, decltype(properties{device_image_scope})> myScopedVar;

class MyKernel;

void device_kernel_function(queue &q, int &output) {
  // Submit a kernel that uses the device_global variables
  q.submit([&](handler &h) {
    h.single_task<MyKernel>([=]() {
      // Access the device_global variables within the kernel
      int val = myGlobalVar;
      float scopedVal = myScopedVar;
      // Perform computations...
      myGlobalVar = val + 1; // Increment the global variable
      myScopedVar = scopedVal + 1.0f; // Increment the scoped variable
    });
  }).wait(); // Wait for the kernel to complete execution

  // Retrieve the computed values from the device_global variables
  output = myGlobalVar;
}

int main() {
  queue q;
  int output = 0;
  device_kernel_function(q, output);

  // Print the output value of the computation occurring in the device kernel
  std::cout << "Output value from device kernel: " << output << std::endl;

  return 0;
}

\end{lstlisting}
\end{figure*}

In the test case code generation stage, the characteristics of the SYCL code that is needed to trigger the function
in the compiler pass is provided along with a query to RAG framework to generate the SYCL code that is used to
test the compiler and trigger the compiler pass function under test. An example of this query is shown in Listing~\ref{query_sycl2}. During the test case code generation phase, the value of the variable {REQ} as shown in Listing~\ref{query_sycl2} represents the
generated characteristic from Stage 1 and incorporates a randomly selected SYCL optimization feature, memory
access technique, and data structures. 

\item \textit{Compilation:} After the code snippet is generated, it is compiled for a GPU target
device. We target GPU devices as they are widely used to accelerate applications implemented in SYCL. If the
compilation fails, the compilation error, the failed code and a query are sent to RAG framework to modify the
code to correct the compilation error. The query used in correcting the code that failed compilation is shown in Listing~\ref{query_sycl3}. This process is repeated up to 5 times if a correctly compiled code is not generated. The generated code
that compiles without errors is then modified $X$ times by random selection of SYCL optimization feature, memory
access technique, data structures, and random number of device kernels. The value of $X$ is chosen by the user
of the tool. To modify the generated code, the query shown in Listing~\ref{query_sycl4} is sent to RAG framework to generate the
modified codes. The generated modified codes are then compiled, and any compilation error is corrected using
query in Listing~\ref{query_sycl3}. Finally, the generated code snippets that are compiled correctly are offloaded to multiple
different architectures from different vendors. 
\item \textit{Testing:} The outcome of the compilation and running the generated
binaries on different architectures is compared. Bugs are detected if the compiled binaries crash or
demonstrate different runtime outputs or errors across different architectures. This framework reveals two
classes of issues, (1) runtime errors or crashes due to a potential issue in the implementation of a compiler
pass on a particular device architecture; and (2) inconsistent runtime behavior of the generated binary across
different device architectures.
\end{enumerate}

\begin{figure*}[t]

\begin{lstlisting}[frame=single,caption={Query that is used to generate testcase code.},label={query_sycl2}]

	You are an expert compiler developer. Please generate a valid device kernel Code implemented in SYCL programming model that meets the requirements below. The code should contain a `main` function. And the main function gives back an output value. Please initialize all the variables you define with a value. The code must comply with sycl specifications and specifications of lowerIR sycl optimization """+ {Name of optimization pass}+"""	
	> Requirements: 
	```
	{REQS}
	```	
	> Generate the final code between the following pattern %%START%% and %%END%%
Include the following header files
   #include <sycl/sycl.hpp>
   
    #include <array>
    
    #include <iostream>
    
   #include <string>
   
    #if FPGA_HARDWARE || FPGA_EMULATOR || FPGA_SIMULATOR
    
    #include <sycl/ext/intel/fpga_extensions.hpp>
    
     #endif
     
     using namespace sycl;
     
    generated code is used by compiler validation team to test compiler
    
    ALL GENERATED CODE MUST BE COMPLETE AND MUST COMPILE CORRECTLY.
    
    Make sure to use namespce sycl::ext::oneapi::experimental for device global.
    
After generating the final code, check by yourself that all code requirements are satisfied and that code is COMPLETE and COMPILABLE
\end{lstlisting}
\end{figure*}

\begin{figure*}[t]
\begin{lstlisting}[frame=single,caption={Query that is used to fix compilation error.},label={query_sycl3}]
When this code {Generated code}+  is compiled using clang++, I got the following compilation error +{ Compilation Error}+ fix this error and generate the full complete code. Include all needed headers and namespaces. Make sure to use namespce sycl::ext::oneapi::experimental for device global. Generate code between the following pattern %%START%% and %%END%% 

\end{lstlisting}
\end{figure*}

\begin{figure*}[t]
\begin{lstlisting}[frame=single,caption={Query that is used to modify the generated code.},label={query_sycl4}]
You are an expert compiler developer. Please modify the code below as per the    requirements below to generate FULL COMPILABLE CODE. 
	> CODE: 
	```
	{CODES}
	```	
	> Requirements: 
	```
	{REQS}
	```
	>  Generate the final code between the following pattern %%START%% and %%END%%
Include the following header files
   #include <sycl/sycl.hpp>
   
    #include <array>
    
    #include <iostream>
    
   #include <string>
   
    #if FPGA_HARDWARE || FPGA_EMULATOR || FPGA_SIMULATOR
    
    #include <sycl/ext/intel/fpga_extensions.hpp>
    
     #endif
     
     using namespace sycl;
     
    generated code is used by compiler validation team to test compiler
    
    ALL GENERATED CODE MUST BE COMPLETE AND MUST COMPILE CORRECTLY.
    
    Make sure to use namespce sycl::ext::oneapi::experimental for device global.

After generating the final code, check by yourself that all code requirements are satisfied and that code is COMPLETE and COMPILABLE
\end{lstlisting}
\end{figure*}

\section{Experimental Setup and Results}
\label{exp_results}
In this section, we describe our experimental setup, issues found with the framework, and provide a cost estimation for the LLM usage.
\subsection{Experimental Setup}
We implement the proposed end-to-end tool in Python 3.10.12. We use OpenAI-Azure GPT4 as the LLM used for generation in the RAG framework. We use azureopenai-text-embedding-ada-002 as the embedding generation engine. We use PostgreSQL as the vector
database. We use the proposed framework to generate code snippets that target five compiler passes in clang++ version 19.0.0git that
supports SYCL implementations and is built from source using the following clone
(https://github.com/intel/llvm.git 0f796bcde4d1815ff990388d0f0112379b9739b0). The binaries generated
from this compiler have been tested on a NVIDIA A40 GPU. We also use the same
generated SYCL code test cases to test Intel(R) oneAPI DPC++/C++ Compiler 2024.2.0 (icpx). The binaries generated
by this compiler are offloaded to the following target devices: (1) Intel(R) Xeon(R) Platinum 8480+ OpenCL 3.0;
and (2) NVIDIA A100 80GB PCIe 8.0. We use the flags \texttt{-fsycl -fsycl-targets=nvptx64-nvidia-cuda} to target
Nvidia GPU devices. We use the following flags: \texttt{spir64, spir64\_x86\_64, spir64-unknown-unknown, spir64\_x86\_64-unknown-unknown flags} to target Intel(R) Xeon(R) Platinum. We compile all source codes with optimization levels -O0, -O1, -O2, -O3 for all the compilers and devices under test. 
We run the script
for code generation for five compiler passes.
The compiler passes that we tested in this work are: (1) Device globals~\cite{deviceg_} which is an optimization feature that
enables the definition of global variables that are shared across device kernels and the host.; (2)
LowerWGScope~\cite{LowerWGScope_} which is responsible for transforming LLVM intermediate representations (IR) to semantics that
are compliant with SYCL specifications when the code is offloaded to different architectures; (3)
LowerWGLocalMemory~\cite{LocalAccessorToSharedMemory_} which lowers the local memory allocations in SYCL kernels; (4)
LocalAccessorToSharedMemory~\cite{LocalAccessorToSharedMemory_} which transforms SYCL local accessors to shared memory allocations in GPUs;
and (5) CompileTimePropertiesPass~\cite{CompileTimePropertiesPass_} which processes compile-time properties of SYCL variables.
\subsection{Issues Detected}
The total number of generated code that correctly compiles and is used to test the five compiler passes is 152
SYCL source files.
We performed differential testing across NVIDIA architectures by executing binaries compiled using Clang++ and icpx with different optimization flags enabled (from the same testcase source code) on NVIDIA A40 and A100, respectively. Thus, we compare the binaries generated from two compilers on two different GPU devices. We observe that 12 files lead to different output values to be observed when different
optimization levels are used to compile source code on icpx and clang++ or lead to different run-time
errors/crashes. We have noticed that issues resulted from the code that was generated using characteristics
extracted from functions that are implemented in DeviceGlobals, LowerWGLocalMemory, LowerWGScope, and
LocalAccessorToSharedMemory compiler passes. We also performed differential testing by running the same binaries on different devices. We observe the differences in execution results when running binaries compiled with icpx on NVIDIA A100 80G PCIe
and Intel(R) Xeon(R) Platinum. We identified 75 files that demonstrated differences in behavior between
running binaries on the devices. The general observations of different execution behavior include binaries crashing on some architectures and
providing correct values on others. Binaries providing different values on different architectures and features
not supported on some devices but supported on others. 

Upon inspecting the source of misbehavior observed, 1 bug has been confirmed and fixed \cite{fix_issue}. The rest of the observed differences are due to undefined behavior in the code generated by the GPT-4. In addition, in the prompts used for code generation, we have instructed the GPT-4 to generate SYCL code with multiple kernels. We have observed that in most cases, GPT-4 generates SYCL code with multiple kernels but without proper synchronization. Thus, in future work, we would focus on developing a module to prevent the generation of code with undefined behavior and we would perform prompt engineering to force LLMs to generate code with synchronization. 

\subsection{Cost Evaluation}
In the test case generation module, a total of 269 files were generated for the compiler passes. There were 152 successful compilations and 117 unsuccessful compilations.
 
For the characteristic’s generation phase, the characteristics needed to invoke a specific function in a compiler
pass is generated once. To test five compiler passes, we processed a total of 27 files. The total number of input
tokens which represents the query to generate the characteristics that is passed to RAG framework and the
context that is retrieved from the vector database in the RAG framework is 77,608 tokens. The total number of
output tokens that represents the number of tokens for the characteristics needed to trigger the functions in
the compiler pass is 16,570. We use GPT4. So, as per the latest GPT4 pricing \cite{pricing_} GPT4 costs $\$30/1M$ input
tokens and $\$60/1M$ output tokens. Thus the estimated cost for the characteristics generation phase is$ (77608
*30/1M) + (16570*60)/1M=\$3.3$.

This cost can go down to \$0.63 if we use the most recent OpenAI GPT4o
model.
For the testcase code generation stage (stage 2), the average number of input tokens to the LLM is 3,321. This
includes the number of tokens in the query that is sent to RAG framework and the context retrieved from the
vector database within the RAG framework that represents relevant chunks in SYCL and compiler
documentation. The average number of output tokens generated by LLMs which represents the number of
tokens of the generated test case code is 638 tokes. Thus, for generating 269 files, a total of almost 893,349
input tokens and 171,622 output tokens are consumed. Thus the estimated cost for test case generation
performed in these experiments is $(893,349*30/1M) + (171,622*60)/1M=\$37$.

As shown in~\cite{pricing_}, the pricing of
new advanced OpenAI models are significantly lower than GPT4. For example, if we use GPT4o, the cost will
reduce to \$7.

The approximate total cost of \$10 for the test case generation phase is negligible compared to
the compensation of an experienced compiler developer.

\section{Limitations and Future Work}
\label{limitations_}
This framework was able to detect misbehaviors in two compilers covering different architectures.
We have noticed that among the 87 examples that were flagged misbehaviors, the categories of detected misbehavior
was similar in many of the files. This suggests that generated test cases are repeatedly triggering similar flows in the compiler.
Thus, in future work, we plan to introduce a metric that measures the variations among the generated test
codes. In addition, we plan to integrate feedback from compiler passes coverage and utilize the
feedback to generate test codes that trigger different variations of compiler misbehavior. Also, we plan to
inspect every generated file and utilize feedback from the instrumented compiler to investigate whether the
generated code covers every generated characteristic during the characteristic’s generation phase and whether
it triggers the target compiler pass.
Compiler fuzzing campaigns typically require thousands of test cases to maximize the number of detected
misbehavior issues. Thus, in future work, we propose to utilize the RAG framework to generate python scripts that
describes code generation policies required to generate code snippets to test each compiler pass. This would
reduce the cost of LLMs. In addition, we can run the generated python script to generate thousands of code
snippets for each compiler pass.
In addition, we plan to craft code generation queries to generate code that specifically reveal security bugs that
can be induced by the compiler features and compiler passes. We also plan to utilize GPT4o within the RAG
framework to improve the quality of the generated test cases and reduce cost. In addition, we have noticed that the majority of the reported issues are due to undefined behavior in the generated code. Thus, we plan to propose a methodology to prevent the generation of code with undefined behavior. 

\section{Conclusion}
\label{conclusions_}
In this work, we presented an end-to-end framework to fuzz Intel DPC++ compilers. The proposed framework
relies on LLM-based RAG framework. It can efficiently generate test cases for multiple compiler passes within
a few hours without the need for prior knowledge about compiler internals. The proposed framework has been
used to test five compiler passes and was able to detect misbehaviors when compiling SYCL source code and
offloading the code to multiple architectures from different vendors. Going forward, we would continue to test another
compiler passes using the proposed tool. We would introduce future modifications to increase the number of
detected misbehavior and address the highlighted limitations in the current work.

\small
\bibliographystyle{IEEEtran}
\bibliography{bare_jrnl.bib}
\newpage
%






\end{document}